# Ranking of Wikipedia articles in search engines revisited: Fair ranking for reasonable quality?

Dirk Lewandowski
Ulrike Spree

Hamburg University of Applied Sciences, Faculty DMI, Department Information, Berliner Tor 5, D—20099 Hamburg, Germany
dirk.lewandowski@haw-hamburg.de, ulrike.spree@haw-hamburg.de

**Abstract**

This paper aims to review the fiercely discussed question of whether the ranking of Wikipedia articles in search engines is justified by the quality of the articles. After an overview of current research on information quality in Wikipedia, a summary of the extended discussion on the quality of encyclopedic entries in general is given. On this basis, a heuristic method for evaluating Wikipedia entries is developed and applied to Wikipedia articles that scored highly in a search engine retrieval effectiveness test and compared with the relevance judgment of jurors. In all search engines tested, Wikipedia results are unanimously judged better by the jurors than other results on the corresponding results position. Relevance judgments often roughly correspond with the results from the heuristic evaluation. Cases in which high relevance judgments are not in accordance with the comparatively low score from the heuristic evaluation are interpreted as an indicator of a high degree of trust in Wikipedia. One of the systemic shortcomings of Wikipedia lies in its necessarily incoherent user model. A further tuning of the suggested criteria catalogue, for instance the different weighing of the supplied criteria, could serve as a starting point for a user model differentiated evaluation of Wikipedia articles. Approved methods of quality evaluation of reference works are applied to Wikipedia articles and integrated with the question of search engine evaluation.

*"Utility ought to be the principal intention of every publication. Wherever this intention does not plainly appear, neither the books nor their authors have the smallest claim to the approbation of mankind." (William Smellie: Preface of the 1st edition of the Encyclopedia Britannica 1768)*

## 1. Introduction

In June 2004 the well-known Austrian blogger Horst Prillinger deplored in a rather emotional contribution the unjustified top ranking of "badly researched, incorrect Wikipedia articles" in search engines (Prillinger, 2004). The study presented in this article systematically reassesses the question of how Wikipedia entries rank among informational queries in popular search engines and whether the quality of the retrieved articles justifies their ranking position. We analyzed the ranking of Wikipedia articles for 40 informational queries in 5 popular search engines (dataset from Lewandowski, 2008).

The queries were phrased in German and the data was collected on the 20th and 21st of January 2007. The evaluated Wikipedia articles are all from the German Wikipedia. In 2007 the German Wikipidia still allowed its users to edit articles freely and displayed the changes immediately. Since 2008 as default value only sighted versions of an article are displayed, however the user still has access to all versions under the version tab (Wikipedia. (2009, July. 28 23:03)).

The quality discussion regarding Wikipedia—not only in the mainstream media—is, as Prillinger's statement illustrates, very much focused on aspects of correctness in form and content (Giles, 2005). This narrowing of the quality debate does not meet the requirements of the complex discussion on lexicographical quality in general. In this discussion, correctness is just one facet in a set of criteria. In this article we shed new light on the question of lexicographic quality as our pragmatic research question is whether the overall quality of the retrieved articles, correctness included, justifies its ranking, or in other words, is the ranking of the articles appropriate to their usefulness for the user of the search engine? Our research is also to be seen in the larger context of adding transparency to the question of how search engines deal with Wikipedia entries. Encyclopedias are generally recommended as a starting point for information research. Does the ranking in Wikipedia articles do this view justice?

To provide some contextual background we summarize briefly the current research on information quality in Wikipedia as well as the extended discussion on the quality of encyclopedic entries in general. On this basis, we develop our own heuristic method for evaluating Wikipedia entries.

## 2. Research objectives and questions

Our actual research covers two closely linked issues. We wish to provide an easy to manage but reasonably reliable tool to evaluate Wikipedia articles and to gain deeper insight into whether user judgment of Wikipedia articles retrieved by search engines is in accordance with independent quality judgment of Wikipedia articles. Our main focus is to gain a better understanding of the interrelatedness of the quality judgment of a Wikipedia article, its ranking in search engines and user judgment of the adequateness of this ranking. Based on the literature review and our results on ranking of Wikipedia articles in search engines, we formulated the following research agenda that will guide us through the discussion of the heuristic evaluation of Wikipedia articles:

1. Which applicable quality standards (heuristics) exist for evaluating Wikipedia articles? In what context were they developed and applied and do they justice to the generic markings of Wikipedia articles?
2. Based on the research on existing quality standards we developed our own heuristics. With the help of these heuristics human evaluators should be able to make sound and intersubjectively comprehensible quality judgments of individual Wikipedia articles. As we wanted to develop an easy to apply tool our heuristic had to meet the following requirements:
    a. Human evaluators can evaluate individual Wikipedia articles on the basis of the provided criteria catalogue and can agree whether a given article meets a certain criterion or not.
    b. On the basis of the criteria catalogue human evaluators attain similar evaluating scores for the same article.
    c. On the basis of the criteria catalogue noticeable differences in quality of Wikipedia articles can be determined.
3. The calibrated heuristic was applied to Wikipedia articles that scored highly in the retrieval test to find out

a. whether there exist noticeable differences in quality among the examples of our sample,
        b. and whether there are really bad articles among the highly ranked articles.
    4. On this basis new insight into the user judgment of Wikipedia hits is possible as it now can be analyzed
        a. how user relevance judgments of the Wikipedia hits in the search engine results correspond with scores from the heuristic evaluation,
        b. how useful the ranked articles are,
        c. and whether the ranking is appropriate, respectively whether good entries are ranked high enough.

### 3. Literature overview

Since its launch in 2001, Wikipedia has become a well-researched object in diverse disciplines, especially in computer science, linguistics and literary studies, and social sciences. Apart from studies that use the large Wikipedia text corpus as a source for experiments in automatic indexing or for the application of knowledge organization features (Voss, 2006) and to better understand semantic Web applications, the collaborative generation of knowledge, and the peculiarities of social networks are the main research questions (Schroer, 2008; Korfiatis, Poulos, & Bokos, 2006). Schlieker (Schlieker, 2005) analyses the process of the collaborative knowledge production as well as the social background and the motives of the contributors for participating in Wikipedia. In this context scholars exploit the huge international Wikipedia community as an example to get deeper insight into the functioning of open source projects (Anthony, Smith, & Williamson, 2005, 2007).

Simultaneously with the rising success of Wikipedia, an intense discussion on quality aspects in the narrower sense began. In the beginning, academic studies were often triggered by more or less sensational warnings in the popular press about the serious threat that Wikipedia posed for professional academic research. Quality deficiencies were above all anticipated in three areas: a main quality concern is the question of correctness in form and content, closely linked to the question of verifiability, followed by discussions on bias and propaganda in Wikipedia, notably for controversial topics, a problem that is closely linked with the whole question of vandalism, and the volatility of content because of the editorial principle of Wikipedia that allows anyone to edit articles freely.

*Correctness in form and content*

A general evaluation of correctness of encyclopedia articles is hardly possible due to the overall scale and the wide range of subject areas, so most of the studies focus on specialized fields of knowledge like science, philosophy, or history und analyze only a small number of articles (Bragues, 2007; Giles, 2005; Rector, 2008; Luyt, 2010). For example, the often-quoted "nature study" (Giles, 2005) that compared articles on scientific topics from Wikipedia and Britannica was based on the analysis of only 42 articles. The much-disputed *Encyclopedia Britannica* (2006) study, however, never convincingly disproved the results of the study indicating that there were no serious differences between the two encyclopedias concerning grave mistakes, whereas more small mistakes were observed in Wikipedia than in *Britannica* (Giles, 2005). Meanwhile, similar results have been obtained in evaluating the quality of health information in Wikipedia (Devgan, Powe, Blakey, & Makary, 2007). To increase the verifiability of Wikipedia, currently several projects are concerned with including different metadata sets and authority files (Danowski & Pfeifer, 2007). There is some evidence with regard to formal accuracy that Wikipedia is less reliable than comparable works (Hammwöhner, 2007; Hammwöhner, Fuchs, Kattenbeck, & Sax, 2007; Schlieker, 2005). However, it can be disputed whether the heuristic assumption that one can easily infer from a high number of orthographical mistakes an equally high number of factual mistakes is valid (Fallis, 2008, p. 1668).

*Vandalism, bias, and propaganda*

Due to its policy that anybody was allowed to edit Wikipedia articles freely and anonymously, Wikipedia was prone to biased editorial changes like glossing over of biographies or maliciously changing information about political adversaries ("Dirty Wikitricks," 2006) as well as the manipulation of articles about companies and organizations by themselves (Ludwig, 2007). Once the changes were detected—in the past the press revealed several such cases of vandalism whereas the dissociation between investigative journalism and muckraking was always fluent (Lischka, Patalong, & Christian, 2007) — they were usually undone immediately. These kinds of changes are rarely outright lies but mostly a glossing over of facts. For example, in an article on Wal-Mart, the phrase "Wages at Wal-Mart are about 20% less than at other retail stores" (2005/05/05 22:21) was changed to "The average wage at Wal-Mart is almost double the federal minimum wage" (2005/05/05 22:25). Changes can be followed easily using the version history of Wikipedia, The German Wikipidia in 2009 introduced the feature *sighted version* as a further protection against vandalism. In the English version this feature is still in the proposal status (2008/05/16 20:16). Unregistered users can still edit articles freely, however the articles are now marked as draft version as long as they have not been checked by a regular Wikipedia author and marked as sighted (2009/07/28 23:03).

Although the academic evaluation of Wikipedia's quality is well-received and a limited number of studies is cited, in most articles there is no consensus on the appropriate methods of evaluating encyclopedias, and the results are hardly comparable as they differ considerably regarding choice and number of evaluated articles, applied quality criteria, and method of evaluation (Hammwöhner, 2007).

As its pure size does not allow the evaluation of Wikipedia as a whole—a problem with respect to the evaluation of reference works in general—every quality evaluation has to determine the choice of articles. We can distinguish studies based on a small number of deliberately chosen articles on specific topics (Bragues, 2007; Mühlhauser & Oser, 2008) from studies that try to define a representative sample of articles (Hammwöhner et al., 2007). In addition, the studies differ on whether they evaluate the encyclopedia as a whole concerning structure, appropriateness of choice of lemmatization, homogeneity of articles, balanced and appropriate length of articles, and cross-referencing or whether the quality of single articles is evaluated. Most studies tend to the latter, whereas structural aspects are studied in a broader context like knowledge organization in encyclopedias (Voss, 2006).

Since the beginning of quality evaluation in Wikipedia, two different trends can be observed. A purely polemical discussion was followed by intellectual expert evaluation according to defined quality heuristics. After a focus on questions of accuracy and correctness the whole range of quality criteria that are traditionally applied to the evaluation of encyclopedias (for an in depth discussion, see the following paragraph) was applied to Wikipedia. Inspired by the quality management process in Wikipedia itself and the intention to develop tools to improve and monitor this quality management process, researchers have strived to derive at a set of quality measures that could be deduced automatically (Stvilia, Twidale, Smith, & Gasser, 2005a). They sought to isolate text parameters that indicate high information quality like size and structure of discussion on articles, scope and frequency of edits, or number of individual authors of entries. Besiki Stvilia (2008) and his research group imbedded their research on IQ metrics in a large, long-term research project aimed at creating a complete workbench for information quality in general.

Despite the aforementioned huge differences in approach and research questions, the academic discussion on quality aspects in Wikipedia has reached some common ground:

It is agreed that Wikipedia articles generally supply reliable and useful information and are a useful auxiliary to contextualize knowledge (Chesney, 2006). The error rate is not considerably higher than in comparable reference works (Giles, 2005; Mühlhauser & Oser, 2008; Bragues, 2007; Devgan et al., 2007). Results concerning the consistency and comprehensiveness of individual articles also generally have been regarded as satisfying (Hammwöhner et al., 2007). However Wikipedia articles are often criticized as complex and difficult to understand (Mühlhauser & Oser, 2008).

Although the danger of vandalism is a continual problem, it is counterbalanced by the well-functioning self-monitoring process of Wikipedia. Apart from the prominent cases of biased changes of controversial Wikipedia entries by companies or individuals, for instance politicians, studies have observed the lack of important information. For example, Mühlhauser and Oser (2008) in their study on evidence-based health information in Wikipedia articles, observed that the quality of information is comparable between Wikipedia and two German health insurance organizations. However, a variety of important criteria were not fulfilled by any of the three providers.

*Excursus: The quality control process in Wikipedia itself*

Any examination of quality aspects in Wikipedia should take into consideration the ongoing quality control process in Wikipedia itself. The internal quality management in Wikipedia is closely linked to a gradual move towards a certain amount of standardization. The Wikipedia community created a whole set of general policies and more advisory guidelines. The overall policy is limited to the three principles of "neutral point of view," "verifiability," and abdication from original research (Wikipedia, 2008/08/04 22:13). Additionally, a collection of style guides is provided that give detailed recommendations on structuring of articles, headings, formatting, and quotations. The discussion on quality management mirrors a constant tightrope walk between the intention to give advice and orientation and the awareness that too-strict regulations might discourage potential contributors, which is expressed very well in the general principle (fifth pillar) that reads, "Wikipedia does not have firm rules" (Wikipedia, 2008/08/07 7:04). Recent studies on the motivations of Wikipedia contributors have confirmed that contributors feel impaired by too-strict regulations (Nov, 2007; Schroer, 2008).

Instead of formulating detailed editing principles, Wikipedia has chosen another way. In 2002, Wikipedia began to nominate articles of seemingly high quality as "featured articles." The informal process of marking very good articles soon became more formalized and now the featured articles have to pass through a peer review process (Stvilia, Twidale, Smith, & Gasser, 2005b). The German equivalent to featured articles is "excellent articles" ("exzellente Artikel"). On 07/15/2008, the German Wikipedia reported 1,397 exzellente Artikel and on 08/11/2008, the English Wikipedia reported 2,175 featured articles (Wikipedia, 2008/07/15 20:15; Wikipedia, 2008/07/29 16:13). The criteria concern the following four aspects: 1. content and writing style with the attributes of being well-written, comprehensive, and factually accurate; 2. neutrality and stableness; 3. the observance of style guidelines like lead, appropriate structure, and consistent citations; and 4. the appropriateness of images and acceptable copyright status, as well as a balanced relation between length and importance of the article (Wikipedia, 2008/07/29 16:13). In a category below the featured articles are the good articles. The main criteria are basically the same; however, the attributes are less specialized and standardized, for example there is less emphasis on stylistic brilliance and correct citations (Wikipedia, 2008/08/08). The German "Projekt Qualitätssicherung" respectively the English "Wikipedia: Clean up" are attempts to organize a continuous revision process (Wikipedia, 2008/08/11 10:53). There exists a whole range of tags that can be used to mark quality problems in Wikipedia. From tagged articles, a list of pages requiring cleanup is automatically generated. Registered as well as unregistered users can then mend these articles. Cleanup is particularly concerned with issues like spelling, grammar, tone, and sourcing.

Another important tool that was tested momentarily with the German Wikipedia is the Wikipedia Flagged revision/Sighted version. The main objective of this instrument is to fight vandalism. Qualified users—every registered user is automatically granted the right to sight 60 days after his first edit and after 300 article edits—are granted permission to sight articles for changes against vandalism. The sighted articles are marked with a symbol depicting an eye. As the flagged revision tool is still being tested, it has no direct impact on the user as presently unregistered as well as registered users are still both offered the latest version of an article; however, it has been discussed whether the default setting for unregistered users should be changed to always displaying the last sighted version, whereas registered users will always see the latest version. Thus, an unregistered user would never directly come across a newly vandalized article (Wikipedia, 2008/08/09 22:08).

To sum up, we can say that there are a considerable number of valid studies that assess Wikipedia articles on various aspects of quality as well as for their general information value (Fallis, 2008) as well as for specialized information needs. Regardless of these generally positive and encouraging results, there remain a few thus far unresolved quality problems:

- The huge quality differences inside Wikipedia (Bragues, 2007). Schulenberg et al. (2007) addressed the fact that the Wikipedia community so far has developed useful tools to support the genesis of articles of outstandingly high quality (featured articles, writing competition) as well as clearing or improving of very bad or harmful content. The remaining problem, thus, is how to warrant a satisfying medium quality.
- The unbalanced choice and comprehensiveness of articles. This refers to the problem that due to the composition of Wikipedia members and their special fields of interest, sometimes popular topics (pop stars, etc.) are treated more extensively than topics of general or scientific relevance (Schulenburg et al., 2007).

Both problems are closely linked to an important peculiarity of research on Wikipedia: we currently lack critical studies from researchers with an actual lexicographical research background, especially in the field of specialized lexicography. In the theory of specialized lexicography, quality management is firmly grounded on the determination of a user structure consisting of the three aspects of user presupposition: degree of expertise such as layperson or expert, user situation referring to the actual usage such as text production or understanding, and user intention, which can widely vary from gathering factual information to background information or references (Geeb, 1998). So far, Wikipedia has no determined user structure and is trying to serve the needs of the general user as well as the expert. Based on this, it could be concluded that quality problems are to be expected, especially for articles in arcane academic areas like mathematics, as the knowledge gap between the general user and the specialist is large. Interestingly enough, Citizendium, the encyclopedia project of Wikipedia cofounder Larry Sanger, in his standards for a good article contemplates this problem when he admitted that "certain topics cannot be treated except by specialists" and revealed that the Citizendium Foundation in the future may start "an encyclopedia aimed specifically for specialists" (Citizendium, 2008). Don Fallis (2008) took this particular feature of Wikipedia into account when he suggested relative epistemic evaluation, meaning that instead of exclusively measuring Wikipedia against traditional encyclopedias it could be useful to compare Wikipedia articles with other information resources users would use if Wikipedia were not available, such as Web sites returned on an information inquiry by a Web search engine (p. 1667).

## 4. A first glance on the prolonged discussion on quality of encyclopedic entries

As we have seen, discussions on quality evaluation of encyclopedias very much have been focused on overall correctness and reliability. In this article, we are pursuing a broader concept of quality of encyclopedia articles by taking into consideration the long tradition of encyclopedias and reference works as a literary genre (Spree, 2000).

The modern understanding of the quality of encyclopedic entries is the result of a long and disputed historical discussion. Lexicographical entries on the term "encyclopedia" mirror this understanding when they reduce the definition to a common denominator, as in the latest edition of the German *Brockhaus*, which defines an encyclopedia as an extensive reference medium whose keywords inform in alphabetical order about all fields of knowledge. Albeit the alphabetical order as main access point to general encyclopedias did not prevail before the first edition of the Encyclopedia Britannica in 1768, Brockhaus highlights this feature and neglects the fact that important encyclopedia projects also developed elaborate systematic and associative knowledge access points (Spree, 2000, p. 30-32; 50-52). The Wikipedia project itself offers a kind of laboratory for different knowledge access points that still need more research (Hammwöhner 2007). In this article we omit the aspect of knowledge organization in an encyclopedia which is an important quality criterion indeed as we focus on the quality of individual articles. In alphabetical encyclopedias according to *Brockhaus*, the articles should be based on academically sound knowledge and should be easily comprehensible by the general reader (*Brockhaus*, 2008). The aspect of correctness and reliability is implied but not specifically mentioned. This is consistent with the historical self-conception of the genre, striving for truth but expressing awareness of its limits. In talking about encyclopedias, we should not forget that according to one of its founding fathers, William Smellie, the famous *Encyclopedia Britannica* started off as a work of "pastepot and scissors" (Kogan, 1958). The categorical dissociation of conventional reference works from Wikipedia (Danowski & Voss, 2005) belies the long tradition of encyclopedic writing as a collaborative endeavor to collect the knowledge of humankind (McArthur, 1986). In addition to the "great" *Britannica's* roots as an early example of cutting and pasting, consider the French *Encyclopédie* as a dissident enterprise of the French enlightenment. On the website of the Swiss research project on the history of encyclopedias, "Allgemeinwissen und Gesellschaft," Paul Michel compiled the multifarious possible functions of encyclopedias through history, stressing that no individual function is limited to a certain historical period. Functions range from intellectual pleasure over encyclopedias as a substitute for an inaccessible library to order and classifying the universe as a protection against contingency. Encyclopedic entries could also be used to increase the esteem of things or persons and as confirmation of conventional wisdom. Although encyclopedias always have claimed to strive for truth, this did not prevent them from frankly spreading a specific ideology or worldview. The editors of encyclopedias often saw themselves as part of a larger movement of social advancement. This could mean the cultural self-assurance of a certain society or class through encyclopedic knowledge as well as detecting inconsistencies by compiling knowledge or serving as a means of popularization of academic knowledge for a larger audience ("Allgemeinwissen und Gesellschaft," 2003). The founders of Wikipedia are obviously well aware of this multifarious tradition, as a remark of Jimmy Wales demonstrates: "Wikipedia is a work in progress. Mistakes are made during the editing process. […] I think people have the wrong idea of how accurate traditional reference works are" (Nasr, 2006). A general remark like this can hardly be falsified; however, neither the *Encyclopedia Britannica* nor the German *Brockhaus* publicly disclose their editing principles and quality criteria. In responding to an e-mail request concerning the editing principles of the *Encyclopedia Britannica*, the former *Britannica* editor Alex Soojung-Kim Pang replied that he could not remember the existence of a written quality policy and that he assumed that the actual principles were based on the implicit knowledge of experienced editors.

Wikipedia is often criticized as less trustworthy than other reference works because the identity of the author is unknown. Don Fallis (2008) called attention to the fact that this not only holds true for Wikipedia but for encyclopedias in general. In general, trust in an encyclopedia is not derived from trust in the authority of a particular author but in the production process as such (p. 1667; Soojung-Kim Pang, 1998). This idea is wonderfully illustrated by a quotation from a wry little stage play that was put on to celebrate the one hundredth birthday of the German *Brockhaus Konversationslexikon* in 1902 indicated that the *Brockhaus* was such a great book that it still sounds correct if you copy from it faultily (*Der Grosse Brockhaus. Schwank in einem Aufzuge.* Theater des Börsenvereins, 1902, p. 28). Every quality discussion regarding encyclopedias should consider that the accusation of offering unreliable instant or superficial knowledge is part of the historical perception of reference works (Spree, 2000, p. 2). This short historical retrospect should suffice to remind us that it is expedient before applying professionally accepted criteria for evaluating encyclopaedias to measure the quality of encyclopedias against their own aims and objectives as well as the user structure and expectations. Because of the relatively open and unspecific form of encyclopedic entries concerning the definition of distinct characteristics as a literary genre, a quality model should consider this overlap with other forms of informational texts.

## 5. Towards a heuristic method for evaluating the quality of encyclopedic entries

The *OECD Glossary of Statistical Terms* describes quality as a "multi-faceted concept" and underlines its dependency on user perspectives, needs, and priorities. According to this concept, quality requirements can vary widely across groups of users (OECD, 2009). We compared four (one German and three American) reputed publications on quality criteria for reference works published between 1994 and 2002 – within this timespan digital versions (on CD or online) gradually gained acceptance. All four agreed on the following four main criteria (Gödert, 1994; Katz, 2002; Kister, 1994; Crawford, 2001):

- scope, referring to the appropriate coverage of topics,
- authority, referring to the reputation of authors, editorial board and publisher,
- currency / recency, referring to the choice of topics and the frequency of updating,
- organization and accessibility, taking into consideration the various methods of knowledge organization like lemmatization and quality of entry points like indeces or tables of content

Katz, Kister and Crawford also include the aspect of objectivity/viewpoint and (writing) style. . Interestingly enough, accuracy is included only by Crawford (2001). The other three publications derive from a library and information science background. Apart from obvious mistakes and formal inaccuracies, within this context it is simply not feasible to check seriously the factual correctness of a complete encyclopedia. Quality criteria are often relative to the reference work evaluated as they are deducted from the aims and objectives of the considered reference work. Library Associations provide reviews of reference resources on a regular basis cf. (CILIP Information Services Group 2009; ALA American Library Association 2010). Again the released quality criteria correspond to the already mentioned criteria (ALA: "work's purpose, authority, scope and content, organization, and format"; CILIP: authority of the work and the quality and kind of articles and entries, accessibility and arrangement of the information, scope and coverage, style, relevance and quality of the illustrations, quality of indexing, adequacy of the bibliographies and references, currency of the information, physical presentation, originality of the work). Although most Review Journals include works of reference into their portfolio the quality judgments are often on a general level and explicitly dependent on the publications purpose. They concentrate on checking the existence of certain features like atlas, time line, daily features and pick out randomly chosen achievements or mistakes (Bibel, 2008). Simultaneously to the rising importance of Internet reference resources an abundance of guidelines on the evaluation of internet resources has been published since 1995 (Smith, 2009). Here we find examples how the criteria are parameterized. One fine example that served as a paradigm for other similar lists are Alastair Smith's "Criteria for evaluation of Internet Information Resources". The listed criteria are explained by exemplary questions that inform the evaluation, however leave very much room for the individual (subjective) decisions of the evaluators and are on a pragmatic 'hands on' level. The criteria "breadth" as one attribute of "scope" is explained by the question: "Are all aspects of the subject covered?" and to check the accuracy of a resource the checking of the resource against other resources is suggested. (Smith, 2005).

The extended discussion on assessment criteria for reference works mirrors the two efforts to be simultaneously general and specific. At the same time, we can observe a striking stability of agreed-upon quality criteria that are applied in the field of library and information science. To sum up we can observe an overall compliance on the main evaluation criteria however authors neither concur in the exact choice of attributes and the applied labels nor in the form of chosen heuristics and metrics, for instance yes no, scale, open questions with no defined metrics. Usually, they aim at presenting a quality model and leave it to the individual reviewer to substantiate the heuristics according to their purpose, for example suitability of a certain reference work for a certain task or institution.

Studies on the quality of Wikipedia entries rely basically on similar sets of criteria that vary only slightly. For an in-depth discussion of intellectual IQ criteria as well as computable IQ metrics, see Stvilia, Twidale, Smith, and Gasser (2005b) and Stvilia (2008)., Within the sighted research literature we could not identify a general agreed set of professionally accepted evaluation criteria and approaches, this applies to Wikipedia as well as to competing products. It was one important result of the debate on the controversial Nature study on Wikipedia that such a consensus still has to be agreed on (Hammwöhner, 2007). In this article we wish to make a contribution to this ongoing process.

For the purpose of this study, a catalogue of quality criteria that could serve as a foundation for intellectual evaluation by experienced human evaluators was developed. Criteria were derived from the whole set of criteria discussed in the above mentioned studies on the evaluation of works of reference as well as from the internal quality criteria established by the encyclopedias themselves. In particular we rely on the requirements for featured and/or good articles in Wikipedia (Wikipedia, 2008/07/29 16:13; Wikipedia, 2008/08/08), and the standards of a good Citizendium article (Citizendium, 2008/05/09).

The decision for a heuristic method that could be applied by human reviewers is consistent with the reviewing of the search engine results that were also evaluated by human reviewers. The chosen IQ measures refer to individual articles, a comparative evaluation of appropriateness, such as in size and comprehensiveness of articles, is not intended. As the heuristic is adapted to human evaluators, the 14 main criteria are further specified by attributes. We tried to define these attributes precisely enough for human evaluators to agree on them—in this respect we built on the principles of heuristic evaluation as they are known from usability evaluation—however, they would not be computable (Mack & Nielsen, 1994). We focused mainly on content quality aspects. Correctness of facts was checked only on a general knowledge level but not validated by domain experts. Altogether in our choice of criteria, we tried to strike a balance between reference work/lexicography-specific criteria and general content criteria as well as the special formal peculiarities of Wikipedia.

**Table 1: List of applied quality evaluation criteria**

1. Labeling/lemmatization — Obvious/non-ambiguous
   Common usage
2. Scope — Stays focused on the topic (W)
   No original research (W)
3. Comprehensiveness — Addresses the major aspects of the topic (W)
   Understandable as independent text
4. Size — Concise (W)
   No longer than 32 KB (W)
   Appropriate to the importance of topic (W)
5. Accuracy — Orthographically and grammatically correct (W)
   Consistency (concerning names, quotes, numbers, etc.) (W)
6. Recency — Up to dateness of cited or recommended resources
   Up to date/developments of the last 3 month are covered
7. Clarity and readability — Definition
   Concise head lead section (W)
   System of hierarchical headings (W)
   Informative headlines (W)
   Factual
   From the specific to the general
   Coherent writing
8. Writing style — News style/summary style (W)
   Formal, dispassionate, impersonal (W)
   Avoiding jargon
   Contextualization
   Concise
   Logical
   Avoiding ambiguities
   Avoiding redundancies
   Descriptive, inspiring/interesting
   Clear/using examples
9. Viewpoint and objectivity — Neutral
   Fair and traceable presentation of controversial views
10. Authority — Verifiable facts (W)
    Reliable sources
    Informative academic writing style
    Longevity/stability
11. Bibliographies — Uniform way of citation (according to style guide)
    Quotations
    Further reading
    External links
12. Access, organization, and accessibility — Internal links
    External links
    Table of contents
13. Additional material — Pictures and graphics
    Self-explanatory images and graphics
    Captions (W)
    Copyright statement
    Special features
    Tabulary overviews
14. Wikipedia ranking — Featured
    Good

(Attributes marked with W are derived from Wikipedia)

Choice and composition of criteria and attributes are based on the assumption that in reference works in general, there is a close interdependency between style and content. For example, "avoiding jargon" refers to questions of style as well as to questions of content.

The criteria catalogue was transferred into an Excel spreadsheet and the attributes were weighted. As a default setting, most attributes were set at 1, with the exception of writing style. To avoid a preponderance of this category, some attributes were set at 0.5. The decision to treat all criteria as equally important by setting them on 1 or 0.5 is arbitrary and has a purely exploratory function. In accordance with our theoretical assumption (see section 4) that the quality of an encyclopedia article should always be evaluated not only against the aims and objectives of the encyclopedia but also against its user structure and expectations we strove to design a flexible and adaptable heuristic. For example in a test scenario where the suitability of Wikipedia articles for academic purposes is evaluated the aspects 10. Authority and 11. Bibliographies could score higher whereas if one wanted to test the appropriateness of Wikipedia to diffuse general knowledge 7. clarity and readability and 8. writing style could be rated higher. An article that fulfilled all requirements could score 48 points. As all criteria are not applicable to all articles, in order to allow a comparison, it is also possible to compare the articles according to the compiled criteria.

## 6. How do users (and search engines) judge Wikipedia articles?

In this section, we give empirical evidence for the preference of Wikipedia results in Web search engines. In addition, we discuss whether Wikipedia results receive their status in search engines deservedly (measured by user judgments) or if it just results from a deliberate choice of the search engines.

Höchstötter and Lewandowski (2009) studied the structures of search engine results pages (SERP), i.e. what elements are presented to the user in the visible area of the SERPs. The visible area is defined as containing all elements that can be seen without scrolling down the page. The visible area depends on the screen size (or window size, respectively) and can contain elements in addition to the organic results such as advertisements and so-called shortcuts (i.e., specially highlighted results that do not come from the general web crawl of the engine). In accordance with earlier research (Nicholson et al., 2006), the authors found that there was limited space for organic results (only four to seven results in the visible area). In addition, they found that this space was often occupied by results from some "preferred" hosts. Preferred in this case must not mean that these results are deliberately preferred (i.e., they are manually boosted into the top results—although this can be the case sometimes ( see Höchstötter & Lewandowski, 2009, p. 1805). A preference for a certain host could occur simply through ranking algorithms relying largely on link analysis (Thelwall, 2004).

Empirical results for 1000 queries were collected from a log file of a major Web search engine, where 500 queries were very popular queries and 500 from the heavy tail, i.e. queries that were seldom posed. The results showed that there were some very popular hosts, and by far the most popular was Wikipedia (followed—with far fewer results—by hosts such as YouTube, Amazon, and the BBC). This held true for all search engines under investigation (Google, Yahoo, MSN/Live, and Ask.com), while the number and distribution of Wikipedia results differed quite a lot.

For the 1000 queries, Yahoo showed the most Wikipedia results within the top 10 lists (446), followed by MSN/Live (387), Google (328), and Ask.com (255). In looking at the distribution of the results, it could be seen that when Google showed Wikipedia results, these were shown significantly more often in the first positions than with any other search engine. The skewed distribution towards the top results also occurred with Yahoo and MSN (but not to such a degree as with Google). The distribution within the results from Ask.com was much wider.

In conclusion, the authors found that "regarding the visible area, one can clearly see, from the data, the reason users might come to the conclusion that search engines in general and Google in particular

'always' show Wikipedia results" (p. 1805). As an example, regarding popular queries, Google showed a Wikipedia result in the assumed visible area of four results for 35 percent of all queries. The results reported here held true for the U.S. sites of the search engines. Because search engines not only provide different results for different countries but also use at least slightly varied results presentations, these results could not be completely extrapolated to other languages. However, we assume that similar host distributions will occur for other languages. To our knowledge, there is no systematic evaluation of the occurrence of Wikipedia results in German Web search engines yet. Höchstötter and Lewandowski's (2009) study showed the distribution of Wikipedia results in the top 10 results of search engines. It would be interesting to see whether this heavy placement is justified from the user's perspective. Therefore, we re-analyzed the results of judgments from our study on search engine retrieval effectiveness (Lewandowski, 2008). The study used 40 informational queries from the general purpose context. Jurors were the same students that were originally asked to provide queries.

For the present article, we re-analyzed the data set using Wikipedia results only. In the original data acquisition, we collected three types of relevance judgments: binary, five-point scale, and 101-point scale. While we used the binary judgments for our general analysis (Lewandowski, 2008), for a more detailed discussion, we will use all three types of scales in the following analysis.

The following analysis is only exploratory in that the number of cases (N=43) is low. However, as can be seen from table 2, the number of cases corresponds roughly to the distribution discussed above. Yahoo and Google showed the most Wikipedia results within its results sets; the numbers for Ask.com and Seekport (not investigated in the study reported above) are lower. MSN was an exception in that the number of Wikipedia results was much lower.

Table 2 shows the relevance judgments for all search engines and all types of judgments (i.e., binary, five-point scale, 101-point scale). We extracted the Wikipedia results from the general retrieval effectiveness study for this purpose.

In general, we see that users give high judgments for the Wikipedia articles with numbers well above the average relevance judgments for top 10 search engine results (cf. Lewandowski 2008, p. 920). As can be also seen from the data, Ask.com received the best user judgments for the Wikipedia results shown. All Wikipedia results received a binary judgment of 1 (is relevant), and the numbers on the 5-point and 101-point scale were also the best compared to those of the other engines. This shows Ask's strategy to only show Wikipedia results when they are surely relevant. This accounts for a lower number of such results.

In comparing Google and Yahoo, the investigation confirmed the results reported above indicating that Yahoo shows more Wikipedia results than Google (in this case, only slightly more). It is interesting to see that depending on the scale used, users judged the quality of the articles differently. Therefore, and because the differences are only small, we do not recognize any discernible differences in showing relevant Wikipedia results with both engines.

It should be kept in mind that the relevance judgments for the Wikipedia articles not only consider the quality of the article itself but also how well the article matches the query. Therefore, the relevance judgments will not directly correspond to the results from the heuristic evaluation given below.

**Table 2:** Relevance judgments for all Wikipedia results

| Search engine | Average relevance judgment | | | Number of Wikipedia |
|---|---|---|---|---|
| | Binary | 5-point scale | 101-point scale | |

|  |  |  |  | results |
|---|---|---|---|---|
| Ask.com | 1.00 | 3.64 | 85.55 | 11 |
| Google | 0.77 | 2.65 | 65.84 | 31 |
| MSN | 0.71 | 2.57 | 63.57 | 7 |
| Seekport | 0.63 | 2.28 | 54.50 | 19 |
| Yahoo | 0.82 | 2.67 | 60.82 | 33 |

It could be assumed that results clustering will lead to irrelevant results. Clustering in this case means that not all results from a certain host will be shown in the results list, but only two results and a link to more results from that server. The second result from the same host is called a "child" (Höchstötter & Lewandowski, 2009, p. 1799) and is usually indented (see Fig. 1). For practical reasons, in table 3, only the judgments for the second Wikipedia results are compared.

Again, the table shows all types of relevance judgments for all the search engines under investigation. The data set this time consist of only the second results from Wikipedia, i.e. the "children".

The results do not confirm that children results are generally irrelevant. However, they are less relevant than the first ones, but not to a degree which one might expect.

However, again the low number of cases should be kept in mind. The data also does not allow the conclusion that clustered second results are as relevant as first results. To test this hypothesis, further research is needed.

**Fig. 1:** Results presentation with indented second result ("child") and link to more results from the same host

**Table 3:** Relevance judgments for clustered Wikipedia results—second result only

| Search engine | Average relevance judgment | | | Number of Wikipedia results |
|---|---|---|---|---|
|  | Binary | 5-point scale | 101-point scale |  |
| Ask.com | 1 | 4 | 100 | 1 |
| Google | 0.83 | 2.5 | 60 | 6 |
| MSN | 0.5 | 2 | 50 | 2 |
| Seekport |  |  |  | 0 |
| Yahoo | 0.5 | 1.75 | 42.5 | 4 |

When looking at the relevance judgments regardless of the providing host, we found that in all search engines, Wikipedia results were judged better than results from other hosts. However, we still need to know whether they are also judged better than results from other hosts when they fall in the same results position.

We compared relevance scores for Wikipedia results to the average relevance score on the rank where the individual Wikipedia result was found. E.g., if the Wikipedia result was found on rank 3, we used the average score for results on rank 3 for our comparison. This shows us whether Wikipedia results are indeed judged better than other results.

The data in table 4 clearly confirms that Wikipedia results are judged much better than the average results at the same ranking position. However, our analysis cannot unequivocally explain the reasons for this. As jurors were well aware of the origin of the results, it could be possible that they were biased in their judgments due to their prior experience with Wikipedia.

The data indicates that contrary to the assumption that Wikipedia articles show up too often in the search engines' results, the search engines could even think of improving their results through providing more Wikipedia results in the top positions. As the data used for the relevance judgments is from 2007 (Lewandowski, 2008) and the data for the host distribution reported above is from 2008 (Höchstötter & Lewandowski, 2009), it could well be that in the meantime search engines reacted to that fact and further boosted Wikipedia results.

**Table 4:** Relevance judgments for Wikipedia results vs. expected relevance scores

|          | Binary    |             | 5-point scale |             | 101-point scale |             |
|----------|-----------|-------------|---------------|-------------|-----------------|-------------|
|          | Wikipedia | all results | Wikipedia     | all results | Wikipedia       | all results |
| Ask      | 1.00      | 0.50        | 3.64          | 1.31        | 85.55           | 28.05       |
| Google   | 0.77      | 0.61        | 2.65          | 1.83        | 65.84           | 42.97       |
| MSN      | 0.71      | 0.35        | 2.57          | 0.97        | 63.57           | 21.68       |
| Seekport | 0.63      | 0.44        | 2.28          | 1.18        | 54.50           | 27.07       |
| Yahoo    | 0.82      | 0.63        | 2.67          | 1.87        | 60.82           | 43.68       |

In conclusion, we can say that it is state of the art with all search engines to prominently show Wikipedia results. This prominence complies with the users' judgments of these results. However, we still do not fully understand the reasons for user preference for Wikipedia articles. To better comprehend user judgments we examined whether they corresponded with the results from an evaluation based on our set of quality criteria for general encyclopedias.

## 7. Methods for the heuristic evaluation of Wikipedia articles

For the evaluation of the articles, we used the criteria described in section 5. A total of 43 articles were analyzed. A pilot test was conducted to see whether the methods chosen were reliable.

### 7.1. Pilot test

For the pilot test, the authors of this article independently assessed the articles "Düsseldorfer Hafen," retrieved on the query "nightlife in Düsseldorf" ("Nachtleben in Düsseldorf"), "Einfaches Leben," which was retrieved on the query "cost of living in USA" ("Lebenshaltungskosten in den USA"), and "Granularsynthese" which was retrieved for the query "granular synthesis." Within the independent assessment we arrived at similar scores. We took this as an indicator that the descriptions of the applied heuristics were precise enough to derive a common understanding. In the preceding retrieval test, evaluators had marked the articles "Einfaches Leben" and "Düsseldorfer Hafen" as irrelevant for the query whereas "Granularsynthese" was marked as relevant. The results of the evaluation did comply with these judgments. "Granularsynthese" was ranked slightly better than the other two examples. As a main result of the pilot test, we concluded that the criteria were intersubjectively comprehensible; a few minor mistakes in the labeling of the criteria and the weighting were removed. However, before applying them, an agreement had to be reached on how to weight the different criteria of the heuristic method.

*7.2. Main test*

For the main evaluation, two jurors independently assessed the quality of the Wikipedia entries. All evaluations were based on the Wikipedia articles as of the date when the relevance judgments described in section 5 were collected. The evaluation was carried out by two last year undergraduate students, a likely target group for general knowledge articles in an encyclopedia. As student assistants they had already some experience in similar coding tasks. Whenever one of the jurors was in doubt how to interpret an attribute and/or whether a criterion applied or not to an evaluated article she would add a comment in an extra row in the prepared spreadsheet. The jurors also evaluated the articles we used in the pilot test and produced very similar results. After an extensive discussion, especially of the cases of doubt and comparison of the results, the two different lists were merged into one common list. Here we resorted to a method widely used in usability evaluations to reach agreement among usability expert evaluators (Mack, R. L. & Nielsen, J, 1994). Concerning the applicability of our evaluation tool, it is implied that despite the written specification of the individual criteria, by explaining attributes, a further verbal agreement of the jurors on a common understanding of how to apply the criteria is recommended.

## 8. Results

In this section, we present the results from the heuristic evaluation. First, we give an overview of the total scores received, and then we discuss the different areas of our heuristic method individually. Finally, we will compare the results from the heuristic evaluation to the results from the relevance study. The complete individual scores for every criterion are shown in Table 5; the presentation of the results will focus on the most remarkable results.

With an average of 30.53, the evaluated Wikipedia articles achieve a good score overall (69 percent of the maximum score. When looking at the distribution of the scores for the individual articles (Fig. 2), we find that the distribution is skewed towards the higher scores. Additionally, still five of the articles examined (11%) fall below 50 percent of the achievable points.

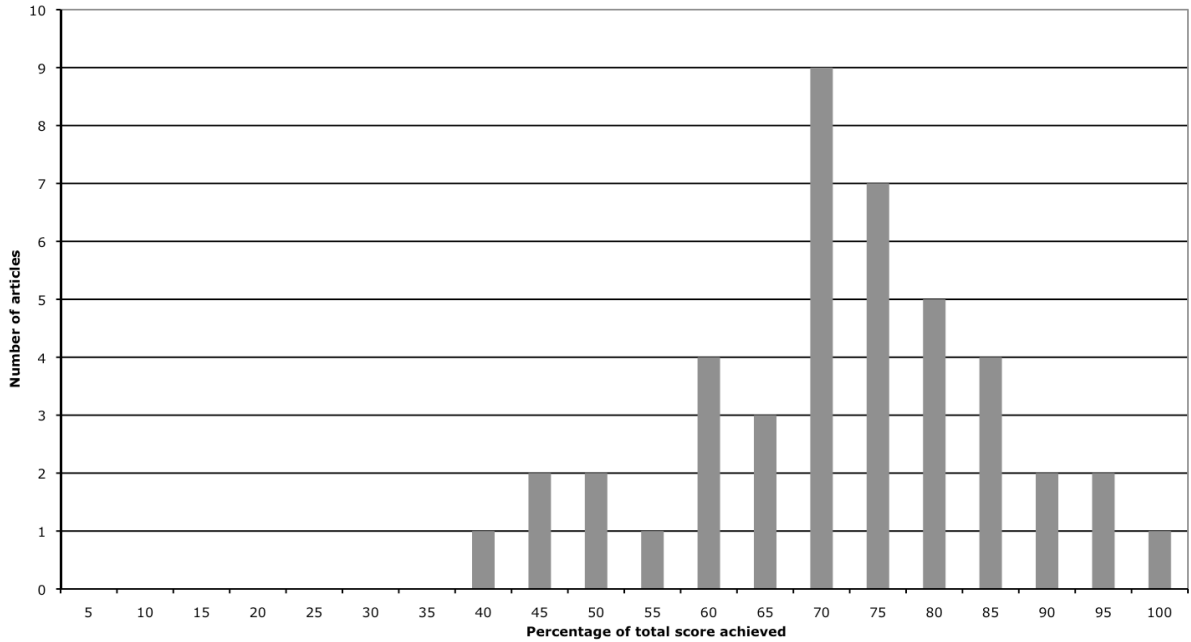

**Fig. 2:** Distribution of overall scores.

*8.1. Scores in the individual criteria*

Table 5 shows the results for the individual criteria. For every criterion, the mean score achieved is given. As the maximum score is 1 in every case, the average score is also the ratio of articles that met the criterion (measured by all articles where the criterion is applicable). This means that for an individual criterion, the score is equal to the percentage of articles where the criterion is met.

The other columns show the percentage of articles that met the criterion, the percentage of articles that did not meet the criterion, and the percentage of articles where the criterion was not applicable. It can be seen that some criteria are not applicable to a relatively large ratio of articles. For example, the criterion "Fair and traceable presentation of controversial views" is not applicable to 60 percent of articles. However, when it is applicable, the criterion is met in 53 percent of cases.

**Table 5:** Overview of article scores.

| | Score (mean) | Criterion met (percent) | Criterion not met (percent) | Criterion not applicable (percent) |
|---|---|---|---|---|
| **Labeling/lemmatization** | **1.81** | | | |
| Obvious/non-ambiguous | 0.89 | 86.05 | 11.63 | 2.33 |
| Common usage | 0.93 | 90.7 | 6.98 | 2.33 |
| **Scope** | **1.74** | | | |
| Stays focused on the topic | 0.81 | 79.07 | 18.6 | 2.33 |
| No original research | 0.95 | 95.35 | 4.65 | 0 |
| **Comprehensiveness** | **1.12** | | | |
| Addresses the major aspects of the topic | 0.29 | 27.91 | 69.77 | 2.33 |
| Understandable as independent text | 0.84 | 83.72 | 16.28 | 0 |
| **Size** | **1.95** | | | |
| Concise | 0.64 | 62.79 | 34.88 | 2.33 |
| No longer than 32 KB | 0.86 | 86.05 | 13.95 | 0 |
| Appropriate to the importance of topic | 0.53 | 46.51 | 41.86 | 11.63 |
| **Accuracy** | **1.40** | | | |
| Orthographically and grammatically correct | 0.67 | 67.44 | 32.56 | 0 |
| Consistency | 0.72 | 72.09 | 27.91 | 0 |
| **Recency** | **0.75** | | | |
| Up to dateness of cited or recommended resources | 0.69 | 55.81 | 25.58 | 18.6 |
| Up to date/developments of the last 3 month are covered | 0.23 | 6.98 | 23.26 | 69.77 |
| **Clarity and readability** | **4.05** | | | |
| Definition | 0.65 | 65.12 | 34.88 | 0 |
| Concise head/lead section | 0.16 | 16.28 | 83.72 | 0 |
| System of hierarchical headings | 0.71 | 55.81 | 23.26 | 20.93 |
| Informative headlines | 0.81 | 69.77 | 16.28 | 13.95 |
| Factual | 0.90 | 88.37 | 9.3 | 2.33 |
| From the general to the specific | 0.66 | 62.79 | 32.56 | 4.65 |
| Coherent writing | 0.48 | 46.51 | 51.16 | 2.33 |
| **Writing style** | **6.49** | | | |
| News style/summary style | 0.61 | 58.14 | 37.21 | 4.65 |
| Formal, dispassionate, impersonal | 0.88 | 88.37 | 11.63 | 0 |
| Avoiding jargon | 0.37 | 37.21 | 62.79 | 0 |
| Contextualization | 0.70 | 69.77 | 30.23 | 0 |
| Concise | 0.59 | 55.81 | 44.19 | 0 |
| Logical | 0.81 | 8.,4 | 18.6 | 0 |
| Avoiding ambiguities | 0.88 | 88.37 | 11.63 | 0 |
| Avoiding redundancies | 0.86 | 86.05 | 13.95 | 0 |
| Descriptive, inspiring/interesting | 0.28 | 18.6 | 48.84 | 32.56 |
| Clear/using examples | 0.65 | 65.12 | 34.88 | 0 |
| **Viewpoint and objectivity** | **1.09** | | | |
| Neutral | 0.90 | 88.37 | 9.3 | 2.33 |
| Fair and traceable presentation of controversial views | 0.53 | 20.93 | 18.6 | 60.47 |
| **Authority** | **3.60** | | | |
| Verifiable facts | 0.95 | 95.35 | 4.65 | 0 |
| Reliable sources | 0.49 | 48.84 | 51.16 | 0 |
| Informative academic writing style | 0.42 | 41.86 | 58.14 | 0 |
| Longevity/stability | 0.93 | 90.7 | 6.98 | 2.33 |
| No obvious mistakes | 0.84 | 83.72 | 16.28 | 0 |
| **Bibliographies** | **2.56** | | | |
| Uniform way of citation | 0.75 | 69.77 | 23.26 | 6.98 |
| Quotations | 0.40 | 39.53 | 60.47 | 0 |
| Further reading | 0.67 | 67.44 | 32.56 | 0 |
| External links | 0.79 | 79.07 | 20.93 | 0 |
| **Access** | **1.70** | | | |
| Internal links | 0.95 | 95.35 | 4.65 | 0 |
| Table of contents | 0.94 | 74.42 | 4.65 | 20.93 |
| **Additional material** | **2.35** | | | |
| Self-explanatory images and graphics | 0.47 | 46.51 | 53.49 | 0 |
| Captions | 0.92 | 53.49 | 4.65 | 41.86 |
| Copyright statement | 1 | 100 | 0 | 0 |
| Special features/tabulary overviews | 0.36 | 34.88 | 62.79 | 2.33 |
| **Wikipedia-ranking** | **0.09** | | | |
| Featured | 0.05 | 4.65 | 95.35 | 0 |
| Good | 0.05 | 4.65 | 95.35 | 0 |
| | | | | |
| **Total score** | **30.53** | | | |
| **Maximum attainable score** | **44** | | | |
| **Attained score (percent)** | **69.16** | | | |

Most articles meet the criteria in the labeling/lemmatization category. The average score for this category is 1.81 (maximum score). Obvious and non-ambiguous labeling applied to 89 percent of the articles considered, with common usage to 93 percent.

Regarding the scope category (maximum score: 2 points), we found that the majority of the articles stayed focused on their topic (81 percent). Also, only five percent of articles contained original research (which is against Wikipedia's guidelines). As can be seen from the data, only 29 percent of the articles addressed the major aspects of the topic. This means that the vast majority of articles left out some of the major aspects, i.e. were incomplete. In this respect, our study confirmed the results of preceding studies regarding Wikipedia's content quality (Fallis, 2008; Mühlhauser & Oser, 2008). Considering the unspecific user model of the Wikipedia that addresses readers at very different levels of expertise, the relatively bad performance of Wikipedia articles in this aspect does not come as a surprise. 84 percent of articles were intelligible as independent texts, while for the rest, additional sources must be considered in order to fully understand the text.

Regarding the size of the articles, the average score was 1.95 out of 3. While most articles were compliant with Wikipedia's allowance of 32 KB, the jurors considered only 64 percent of the articles to be concise. The results were even worse for the appropriateness to the importance of the topic. Here, more than every second article (53 percent) was considered not to be of appropriate length. This could be interpreted as a confirmation of the widely bemoaned observation that Wikipedia is biased towards topics of popular culture; however, our sample is too small and the reasons for inappropriate length of articles too varied for secure validation.

Two thirds of the articles were orthographically and grammatically correct, and 72 percent were consistent concerning names, quotes, and numbers. This surely leaves room for improvement.

Surprisingly, only 23 percent of the articles are up to date, i.e. the developments of the last three months are covered. However, since this criterion was only applicable to a total of 13 articles; results should only be regarded as exploratory.

Regarding the cited sources, the situation was better by far. 69 percent of all articles cited or recommended sources that are up to date.

In the clarity and readability section, the two most obvious results were the lack of a concise head/lead section in most articles and the low ratio of articles written coherently. There was obviously a problem with the head sections of the articles analyzed. Only 16 percent of articles provided a concise head/lead section. This could result from the relatively demanding task of creating such concise head sections. However, as this criterion was derived from Wikipedia itself, it comes as a surprise that Wikipedia does not apply better editorial control with regard to this feature. However, it was counterbalanced by the high score for definitions. It could well be the case that authors eschewed the task of providing concise head sections as well as consistent definitions.

In only 48 percent of cases, our jurors found the articles to be coherently written. As Wikipedia is a collaborative project, it is not surprising that the articles contain a mixture of styles. However, this results in lower readability.

The other factors in the clarity and readability section (definition, headings, informative headings, factual, from the general to the specific) were each met by at least two thirds of the articles considered. In average, an article received 4.05 points out of seven in this section.

The writing style section is the one containing the most criteria (10). The average article received a score of 6.49 here. The two criteria that received by far the lowest scores were "avoiding jargon" and "descriptive, inspiring/interesting" (37 and 28 percent, respectively). This comes as no surprise, as these style criteria are usually associated with professional writers. This result also is an indication of the necessity for more comparative research between the national/language versions of Wikipedia. The observed lack of inspiring/interesting writing is often attributed to the German academic style. A random comparison with the English version of individual articles seems to support this interpretation.

The criterion of fair and traceable presentation of controversial views applied to only 40 percent of the articles. With the rest, there is no controversy (or at least none known to the jurors). Where applicable, 53 percent of articles treated the controversy fair and traceable. However, that also means that in 47 percent of cases, the articles omitted a controversy or were biased towards a certain position of argument.

In the authority section, articles on average received a score of 3.6 out of 5. The vast majority of articles were stable, presented verifiable facts, and contained no obvious mistakes. However, approximately only half of the articles were based on reliable sources (or may have been based on such sources but did not make that clear to the reader). Only 42 percent of the articles were written in an informative, academic writing style. Again, this is no surprise for a collaborative project and corresponds with the results for coherent writing given above.

When looking at the bibliographies, one finds that more than two thirds of the articles had a further reading and/or an external links section. Citations were uniform in three quarters of the articles.

Nearly all articles had internal links to other Wikipedia entries and provided a table of contents. This allows for easy navigation within the article as well as to further articles giving detail on a topic mentioned in the original article.

Additional material such as pictures and tables was sometimes given. However, less than half of the images and graphics were self-explanatory.

Wikipedia's own awards for good articles (divided into "featured" and "good" articles) only applied to a minority of articles. In our data set, only around ten percent received such an award, where there was the same amount for each type of award.

*8.2. Comparison of scores from the heuristic evaluation with relevance judgments*

Again, it should be stressed that the results from the heuristic evaluation cannot be directly compared to the results from the relevance study. Relevance scores are given not only on the quality of the articles themselves but also on how well they matched the query. As relevance is a multifaceted concept not consistently defined (Borlund, 2003; Mizzaro, 1997; Saracevic, 2007a, 2007b), we cannot derive a pure "article quality factor" from the results.

From the scatterplot in Fig. 3, we can see that relevance judgments often roughly corresponded with the results from the heuristic evaluation, mainly when an article was judged as highly relevant. However, there are some cases in which high relevance judgment was not in accordance with the comparatively low score from the heuristic evaluation. This could result from a high degree of trust in Wikipedia. We can assume that when a Wikipedia article is found by a search engine and found on-topic by a user, he would judge it relevant because of his experience with Wikipedia articles in general. The diagram also shows that all Wikipedia articles received evaluation scores above 50 percent. This could have to do with a floor effect in the evaluation, i.e. even the articles that were judged worst received a relatively good overall score.

Another peculiarity is the case of articles getting high scores in the heuristic evaluation while getting a relevance score of 0. We assume that this simply results from articles that were found off-topic for a certain search query, independent of their inherent quality.

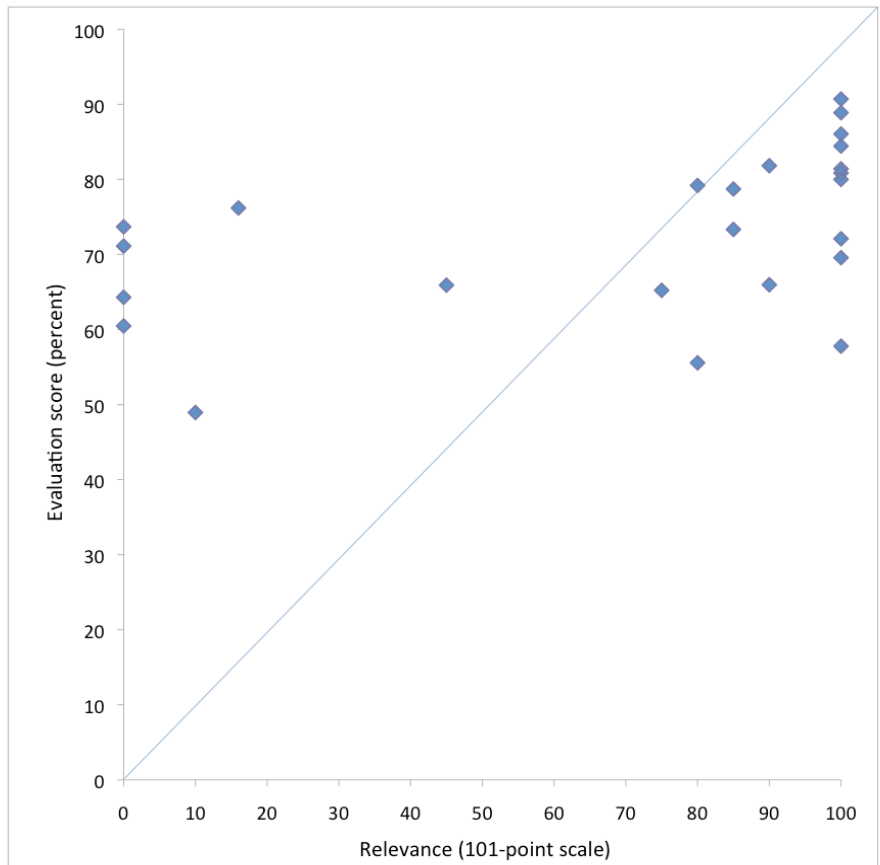

**Fig. 3**: Scatterplot of relevance judgements and results from the heuristic evaulation.

## 9. Discussion and conclusion

In general, our study could confirm that the ranking of Wikipedia articles in search engines is justified by a satisfactory overall quality of the articles. For general informational queries, the negative assessment of Wikipedia articles could not be reinforced with the exception of relatively poor quality concerning orthographical and grammatical correctness.

Our study showed that despite the intense research on Wikipedia quality there is still a lack of commonly agreed on authoritative heuristics as well as evaluation methods (research question 1). However, from the range of existing quality criteria we were able to derive a heuristics adequate for evaluating Wikipedia articles (research question 2). Jurors agreed on the provided criteria catalogue (research question 2 a).

Our heuristic method is apt for the task of detecting quality distinctions, as the quality differences between articles in the sample were clearly noticeable (research question 2 c).

In answer to research question 4b, 4c ("Is the ranking appropriate? Are good entries ranked high enough?"), we can say that the rankings in search engines are at least appropriate. According to the user judgment of relevancy, the search engine providers would even be well-advised to rank Wikipedia articles even higher than they do now.

However, a definite assessment is difficult, as relevance judgment is too multifarious and not solely dependent on content quality of the result. Regarding the correspondence of relevance judgments and scores from the heuristic evaluation (research question 4a), we found some conformance, but as relevance is a multifaceted concept, the results can only give an indication with regard to the reliability of the ranking.

The overall positive impression changes slightly as soon as one considers (expert) queries where more thorough information is needed. For this type of question, the unspecific user model corresponding with the finding that a majority of articles did not address the major aspects of the topic causes considerable problems. On a very general level, it could even be argued that a collaborative approach in knowledge generation results in mediocre/fair/average quality. It can be considered suitable to avoid really bad quality—all of the highly ranked articles scored more than 50 percent—however, very good articles, as in Wikipedia in general, are also an exception. In conclusion, the ranked articles were useful (research question 4b). We did not find articles that were useless (research question 4). However, usefulness varied considerably. While we assume that users' trust in Wikipedia lets them judge most articles as relevant, based on the heuristic evaluation, we cannot recommend always showing Wikipedia results on top of the results list.

Concerning our evaluation tool, the appropriateness of conventional quality critera for evaluating Wikipedia articles can be acknowledged. We revealed noticeable differences in quality among the examples in the sample, as the scores fluctuated between 40 and 95 percent. The intellectual evaluation of articles supported by a standardized tool proved a reliable instrument to measure quality. The only drawback is that this form of evaluation involves a considerable expenditure and is only practicable for a limited sample of articles.

Furthermore, our tool should be seen in the broader context of information literacy education. We developed the prototype of a simple tool that allows users to evaluate individual Wikipedia articles according to their own needs and purposes. This can be seen in the larger context of a necessary change of attitudes towards information quality with respect to Internet sources and especially "participatory encyclopedias" (Haider, 2010). Instead of a stable set of accepted criteria a more flexible set that is adaptable to the user's needs and that adds to making quality judgments obvious and traceable is called for.

It can even be expected that approaches to assess quality of Wikipedia articles automatically could benefit from the results of a qualitative evaluation, as the results could indicate clues for automatically deducible criteria. Just to give an example, if a descriptive/interesting/inspiring writing style coincides with writing practice and training, the experience of an author could be a quality indicator.

In the theoretical introduction to this article, we argued that one of the systemic shortcomings of Wikipedia lies in its necessarily incoherent user model. A further tuning of our standard criteria catalogue, e. g. the different weighing of the supplied criteria could serve as starting point for a user model differentiated evaluation.

**Acknowledgements**

We would like to thank Rita Strebe and Friederike Kerkmann for their help in collecting the data.